\documentclass[journal]{IEEEtran}
\usepackage{amsthm,amssymb,amsbsy,amsmath,dsfont,latexsym,dsfont,array,layout,graphicx,mathrsfs,color,txfonts,amsfonts,bm,color}
\begin{document}

\title{Enhanced Feedback Iterative Decoding \\ of Sparse Quantum Codes}
\author{Yun-Jiang~Wang,
        Barry~C.~Sanders,
        Bao-Ming~Bai,~\IEEEmembership{Member,~IEEE,} and~Xin-Mei~Wang,~\IEEEmembership{Member,~IEEE}
\thanks{Yun-Jiang~Wang is with the State Key Laboratory of Integrated Services Network,
Xidian University, Xi'an, Shannxi 710071, P.~R.~China and with the
Institute for Quantum Information Science at the University of
Calgary, Alberta T2N 1N4, Canada
(e-mail: yunjiang.w@gmail.com).}%
\thanks{Barry~C.~Sanders is with the Institute for Quantum Information Science at the University
of Calgary, Alberta T2N 1N4, Canada.}%
\thanks{Bao-Ming~Bai and Xin-Mei~Wang are with the State Key Laboratory of Integrated Services Network, Xidian University, Xi'an, Shannxi 710071,
P.~R.~China.}
\thanks{Manuscript received Month ??, 2009; revised Month ??, 2009.}}

\markboth{IEEE Transactions on Information Theory~Vol.~?, No.~?, Month~2009}%
{Shell \MakeLowercase{\textit{et al.}}: Bare Demo of IEEEtran.cls
for Journals} \maketitle

\begin{abstract}
Decoding sparse quantum codes can be accomplished by syndrome-based
decoding using a belief propagation (BP) algorithm. We significantly
improve this decoding scheme by developing a new feedback adjustment
strategy for the standard BP algorithm. In our feedback procedure,
we exploit much of the information from stabilizers, not just the
syndrome but also the values of the frustrated checks on individual
qubits of the code and the channel model. Furthermore we show that
our decoding algorithm is superior to belief propagation algorithms
using only the syndrome in the feedback procedure for all cases of
the depolarizing channel. Our algorithm does not increase the
measurement overhead compared to the previous method, as the extra
information comes for free from the requisite stabilizer
measurements.
\end{abstract}


\IEEEpeerreviewmaketitle

\section{Introduction}

\IEEEPARstart{C}{ommunication} is limited by noise in channels, but
error correction methods can efficiently offset this restriction in
both classical~\cite{Gal62,M99} and
quantum~\cite{Sho95,Ste96,Got97,BDSW96,KL97} cases. At a simple
level, multiple copies of the information can be transmitted, and a
majority rule can be applied to discern the correct code, but such
coding is neither practical nor efficient. Sparse graph coding, such
as Gallager's low-density parity-check~(LDPC) codes, offers an
efficient alternative that approaches the Shannon information
limit~\cite{Gal62,M99}. Fortunately quantum coding and decoding
strategies can be constructed from their classical counterparts, but
unfortunately this mapping from classical to quantum coding can be
problematic due to the requirement that quantum codes satisfy the
duality-containing condition~\cite{CRSS98,MMM04}. Moreover, due to
increased challenges posed by these quantum codes, performance
improvement requires further progress in the proposed decoding
algorithm~\cite{PC08,PTO09}.

The entanglement-assisted~(EA) stabilizer formalism adds error-free
entangled bits as a consumable resource for performing quantum error
correction. This EA approach overcomes the duality-containing
requirement and thus offers a rich lode of quantum error correction
protocols inspired by classical protocols~\cite{BDH06}. Using the EA
approach, modern codes for classical channels, such as sparse codes,
can easily be imported as quantum error-correcting
codes~\cite{HBD09,YBZW09}.

Our aim is to improve belief propagation~(BP) decoding methods so
that quantum coding is dramatically improved over existing
techniques. Specifically our numerical results presented here show
that an improved BP method whose heuristical feedback strategies
based on exploiting all accessible information from stabilizer
measurements, yield a dramatically improved block error rate~(BER)
for any depolarizing channel. Our methods should work for any Pauli
noise channel.

For~$n$ qubits, a Pauli channel is defined by the mapping
\begin{equation}
\label{eq:paulinoisechannel}
    \mathcal{B}\left(\mathbb{C}^{2^n}\right)\ni\rho
        \mapsto \sum_{E\in{G_n}}p_EE\rho E^\dagger,\;
    p_E\ge 0, \;
    \sum p_E=1
\end{equation}
with~$G_n$ the $n$-fold tensor product of single-qubit Pauli
operators~$\Sigma\in\left\{I,X,Z,Y=XZ\right\}$. Our interest is
focused on memory-less channels wherein the error on each qubit is
independent of the error on any other qubit. In particular we
consider the depolarizing channel, which is the most-studied
case~\cite{MMM04,PC08,PTO09,HBD09,HI07}: for fixed channel error
probability~$p$, the error on qubit~$q$ is given by
\begin{equation}
\label{eq:depolarizingchannel}
    \mathcal{B}\left(\mathbb{C}^2\right)\ni\rho_q\mapsto p_I
    \rho_q+(1-p_I)\left(X\rho_qX+Y\rho_qY+Z\rho_qZ\right)/3
\end{equation} with $p_I:=1-p$ denoting the probability of no error occurring.

In quantum settings, due to the inability to measure each and every
qubit, syndrome-based decoding is typically chosen. Consequently,
decoders using Pauli channels are generally considered to be
hard-decision decoders. In other words, in quantum decoding, the
conventional soft-decision techniques are not applicable when the
channel is in the Pauli channel model. Generally, sparse quantum
codes are decoded by using syndrome-based BP decoding algorithms
which automatically imply hard-decision decoding. Though there is an
equivalence between syndrome-based decoding and a posteriori
probability decoding (signal-based decoding) under this setting, the
syndrome-based decoding results in a serious drawback to BP decoder:
the symmetric degeneracy error~\cite{PC08}. Fortunately, Poulin and
Chung~(PC08)~\cite{PC08} propose a solution to the symmetric
degeneracy error by using the random perturbation method.

Motivated by the soft-decision techniques used in classical
settings, which provide extra reliable information on the message
nodes thereby yielding a better error correcting
capacity~\cite{M05}, we develop a new heuristical feedback
adjustment strategy for the standard BP decoder. When used in
decoding sparse quantum codes, our method can on one hand solve the
symmetric degeneracy error problem, and on the other hand can
provide more useful information to the message nodes.

The difference between PC08 and our approach is that, in PC08, they
feed back only the syndrome of the decoder output to adjust prior
error probability distributions for received qubits. These adjusted
distributions are then fed back into the decoder. We significantly
improve this protocol by feeding back not just the syndrome but also
the values of the frustrated checks on individual qubits of the code
and the channel model; accordingly we introduce a new adjustment
strategy. Specifically, our approach, which is based not only on
syndromes but also on frustrated checks obtained from full
stabilizer measurements and the channel model, yields a better BER
for the case of depolarizing quantum channels.

We provide a detailed description of our basic BP decoder for decoding
sparse quantum codes. The basic BP decoder
introduced here is inspired by the strategy used for decoding sparse
classical quaternary codes under the BP algorithm. Using this
strategy, we can decode sparse quantum codes directly regardless of whether they arise
from classical binary codes or not and regardless of whether they are Calderbank-Shor-Steane (CSS)
construction~\cite{Ste96,CS96} codes or not.

\section{BP Iterative Decoding}
\label{sec:BPiterative}  In this section, we briefly reprise the
essential elements of BP iterative decoding for sparse quantum
codes. In Subsec.~\ref{subsec:BPdecoding} we discuss the key idea of standard BP
iterative algorithms. Then we compare decoding of classical codes vs
quantum codes, and introduce the standard BP decoding for quantum
codes. Finally in Subsec.~\ref{subsec:decodesparse} we show how to decode sparse quantum
codes in~$GF(4)$ based on the standard BP algorithm.

\subsection{BP decoding algorithm for sparse quantum codes}
\label{subsec:BPdecoding} Consider a $k$-bit message that is encoded
into an $n$-bit codeword, which is then transmitted through a noisy
channel. The received message is  an error-prone vector over the
output alphabet, and there is no guarantee that decoding will reveal
the original codeword. However, the codeword can be guessed with a
high probability of being correct by maximizing the probability for
this codeword based on the observed output vector~\cite{Sho03}.
Unfortunately for a linear block code, encoding~$k$ information bits
into $n$ bits allows $2^k$ possible $n$-bit codewords, and
calculating conditional probabilities for individual codewords is
too expensive in practice.

BP algorithms overcome this inefficiency for sparse
codes~\cite{Gal62}. The strategy is to represent a linear-block
classical error-correcting code (CECC) by a Tanner graph comprising
message nodes and check nodes corresponding respectively to received
bits and check constraints. Then an iterative algorithm recovers the
values of received bits as follows~\cite{Sho03,Mac99}. At each
round, probabilities are passed from message nodes~$\{v\}$ to check
nodes $\{c\}$ and then from $\{c\}$ back to $\{v\}$. The
probabilities from $\{v\}$ to $\{c\}$ are computed based on the
observed value of the message node and other probabilities passed
from the neighboring check nodes to the respective message node. It
turns out that, for decoding sparse codes, the BP algorithm can
provide a reasonable trade-off between complexity and performance.

For a linear block CECC, the code space can be viewed as the
orthogonal projection space~(solution space) of its check matrix.
The sender Alice transmits her message as a codeword that is encoded
according to a specific check~(or generator) matrix through the
channel. When receiver Bob obtains the channel output, the received
vector may not be the solution vector of the check matrix. Therefore
Bob needs to apply a smart algorithm to recover the codeword
efficiently. As the check matrix is sparse, Bob employs the BP
algorithm. First, Bob measures each bit to obtain its posterior
probability distribution. Subsequently, he puts these probabilities
into the BP decoder, and, based on the constraint that the codeword
should be the orthogonal vector of the check matrix, he infers the
original message. This procedure extends naturally to sparse quantum
codes.

In the quantum case, the stabilizer formalism for a quantum
error-correcting code~(QECC) is useful. The code space within
$\mathbb {C}^{2^{n}}$ corresponds to the simultaneous $+1$
eigenspace of all~$m$ generators of an abelian subgroup~\cite{Got97}
\begin{equation}
\label{eq:stabilizer}
    S=\{S_1,\ldots,S_m\}\subset G_n.
\end{equation}
Alice transmits her message as a codeword, which propagates through
the depolarizing quantum channel. Due to channel errors, the
received codeword may not be the simultaneous $+1$ eigenspace
of~$S$. Bob measures~$S$ to obtain the syndrome\footnote{If~$[S_{c},
E]=0$, $s_{c}=1$ or else $s_{c}=-1$, where $c=1,\cdots,m$
and~$S_{\rm c}$ can be viewed as a check node in the Tanner graph of
a QECC with stabilizer~$S$ from Eq.~(\ref{eq:stabilizer}).}
$s=(s_1,\ldots,s_m)\in \{-1,1\}$ and then estimates which error~$E$
from Eq.~(\ref{eq:paulinoisechannel}) has occurred based on the
noise model and the syndrome, which is effectively a check
constraint. Bob redresses the error by applying the same error
operation he inferred onto his received codeword to try to recover
the quantum code. For Pauli channels all errors are $n$-fold tensor
products of single-qubit Pauli operators hence square to the
identity; thus re-applying the same error should restore the
original codeword.

Due to the close analogy between CECCs and QECCs as discussed above,
it is tempting to extend the BP iterative decoding algorithm to
quantum settings. In fact a QECC can also be represented by a
decorated Tanner graph~\cite{COT07,LP08}. Therefore, it is natural
to transmit the messages from $\{c\}$ to $\{v\}$ and the
messages from the $\{v\}$ to $\{c\}$ according to the
principles of BP algorithms~\cite{COT07,LP08}.

Here we first briefly review the BP algorithm used for decoding
sparse quantum codes~\cite{PC08} but begin with necessary
terminology. The probability of Pauli error~$E_q$ occurring on
qubit~$q$ is $p_q(E_q)$. The neighborhood of qubit~$q$ is denoted
by~$n(q)$, and we define $n(c)$ similarly as the set of qubits
connected to check $S_{\rm c}$. The tensor product of Pauli errors
over~$n(c)$ is denoted~$E_{c}$. Then, following PC08's notation, for
two $1$-qubit case, $E \cdot E'=1$ means $[E,E']=0$ and~$E \cdot
E'=-1$ means $\{E,E'\}=0$. For two $n$-qubit Pauli operators such as
$E=E_1\otimes E_2\otimes \dots \otimes E_n$ and $E'=E'_1\otimes
E'_2\otimes \dots \otimes E'_n$, we have
\begin{equation} \label{eq:Pauliproduct}
    E\cdot E' \coloneqq \prod_{k=1}^{n} E_k\cdot E'_k.
\end{equation}

The messages from check nodes to qubit nodes are denoted
\begin{equation}
\label{eq:BPequation1}
    m_{c \to q}(E_q) \propto \sum_{\substack{E_{q'} \\ q'\in n(c)\backslash q}}
        \Big( \delta_{s_{\rm c},S_{\rm c} \cdot E_{\rm c}}
        \prod_{q'\in n(c)\backslash q}m_{q' \to c}(E_{q'})\Big),
\end{equation}
which is defined only up to a constant factor.
The factor can be fixed by normalization~\cite{PC08}.
The messages from qubit nodes to check nodes are similarly denoted as
\begin{equation}
\label{eq:BPequation2}
   m_{q \to c}(E_q) \propto p_q(E_q) \prod_{c' \in n(q) \backslash c} m_{c' \to
   q}(E_q),
\end{equation}
where $p_q(E_q)$ is the initial probability in the definition of the memoryless channel.
Then the beliefs~$b_q(E_q)$ are constructed by first initializing~$m_{q \to c}(E_q) = p_q(E_q)$,
then evaluating according to
\begin{equation}
\label{eq:beliefs} b_q(E_q)=p_q(E_q) \prod_{c\in n(q)}m_{c \to
q}(E_q),
\end{equation}
after the iteration procedure based on Eqs.~(\ref{eq:BPequation1})
and (\ref{eq:BPequation2}). In fact, Eqs.~(\ref{eq:BPequation1}) and
(\ref{eq:BPequation2}) define a sum-product iterative procedure for
decoding sparse quantum codes. Hence this algorithm is also called
the sum-product algorithm (SPA), which is one of the most important
algorithms based on BP.

In order to clarify the BP decoding algorithm for
sparse quantum codes, we now show how to implement
Eqs.~(\ref{eq:BPequation1}) and (\ref{eq:BPequation2}) in GF(4).

\subsection{Decoding sparse quantum codes in GF(4)}
\label{subsec:decodesparse}

There is a convenient isomorphism between the Pauli group $G_1$
generated by~$\{I, X, Z, Y=XZ\}$ and the Galois Field~$GF(4)$ generated by
$\{0,1,\omega,\bar{\omega}=\omega^{2}\}$.
The isomorphism is explained by the element identification
\begin{equation}
    I\leftrightarrow 0,\,X \leftrightarrow 1,\
    ,Z \leftrightarrow\omega,\,
    Y \leftrightarrow \bar{\omega}(\omega^2)
\end{equation}
and the operation identification
\emph{multiplication}$\leftrightarrow$\emph{addition} and
\emph{commutativity}$\leftrightarrow$\emph{trace inner
product}~\cite{Got97}. For the one-qubit case, we have,
\begin{align}\label{eq:isomorphic1}
    P, Q &\in G_1 \leftrightarrow \hat{P}, \hat{Q} \in {\rm GF}(4), \nonumber \\
    [ P,Q ]&=0 \Leftrightarrow {\rm Tr}(\hat{P} \times \bar{\hat{Q}})
    =0, \;
    \{ P,Q \}=0 \Leftrightarrow {\rm Tr}(\hat{P} \times \bar{\hat{Q}}) =1
\end{align}
The isomorphism is readily extended to the $n$-qubit case:
\begin{align}\label{eq:isomorphic2}
    P, Q &\in G_n \leftrightarrow \bm{u}_P, \bm{v}_Q \in {\rm GF}(4)^n, \nonumber \\
    [ P,Q ]&=0 \Leftrightarrow {\rm Tr}(\bm{u}_P \cdot \bm{v}_Q)
    =0,\;
    \{ P,Q \}=0 \Leftrightarrow {\rm Tr}(\bm{u}_P \cdot \bm{v}_Q) =1.
\end{align} where `$\cdot$' used here (between two vectors) is a
regular inner product. That is, for $\bm{u}=(u_1, u_2,\cdots,u_n)$
and ${\bm{v}}=({v}_1,{v}_2,\cdots,{v}_n)$, we have,
\begin{equation}
\bm{u}\cdot {\bm{v}}\coloneqq \sum_{k=1}^{k=n}  u_1\times \bar
v_{1}.
\end{equation}
The addition and multiplication rules of~$GF(4)$ are shown in
Tables~\ref{table:isomorphictable1}
and~\ref{table:isomorphictable2}, respectively.

\begin{table}[htp]
\renewcommand{\arraystretch}{1.3}
\caption{Addition of {\rm GF}(4)} \label{table:isomorphictable1}
\centering
\begin{tabular}{c|cccc}
    $+$&$0$&$1$&$\omega$&$\bar{\omega}$\\ \hline
    $0$&$0$&$1$&$\omega$&$\bar{\omega}$\\
    $1$&$1$&$0$&$\bar{\omega}$&$\omega$\\
    $\omega$&$\omega$&$\bar{\omega}$&$0$&$1$\\
    $\bar{\omega}$&$\bar{\omega}$&$\omega$&$1$&$0$
 \end{tabular}
\end{table}

\begin{table}[htp]
\renewcommand{\arraystretch}{1.3}
\caption{multiplication of {\rm GF}(4)}
\label{table:isomorphictable2} \centering
\begin{tabular}{c|cccc}
    $\times$&$0$&$1$&$\omega$&$\bar{\omega}$\\ \hline
    $0$&$0$&$0$&$0$&$0$\\
    $1$&$0$&$1$&$\omega$&$\bar{\omega}$\\
    $\omega$&$0$&$\omega$&$\bar{\omega}$&$1$\\
    $\bar{\omega}$&$0$&$\bar{\omega}$&$1$&$\omega$
 \end{tabular}
\end{table}

We can import the strategy used for decoding sparse classical
quaternary codes under BP to use for decoding sparse quantum codes.
This adaptation to quantum codes is achieved by transforming the
check criterion~(\ref{eq:BPequation1}) from commutativity to trace
inner product.

For example, suppose $s_c=1$. Then the check criterion for this
check node should be $S_c \cdot E_c=1$, which is equivalent to
$[S_c, E_c]=0$. According to isomorphism~(\ref{eq:isomorphic2}),
${\rm Tr}(\bm u_{E_c} \cdot \bm {v}_{S_c})=0$, which implies that
either~$\bm u_{E_c}\cdot \bm v_{S_c}=0$ or $\bm u_{ E_c}\cdot\bm
v_{S_c}=1$ because ${\rm Tr}(0)=0$ and~${\rm Tr}(1)=0$. We define
$p_q[x]$ as the probability for~$\hat{E_q} \times
\hat{\bar{S}}_{c_q}$ to take the value~$x$, with~$S_{c_q}$ the
$q^{\rm th}$ entry of~$S_c$,\footnote{For $\hat{E_q}$ and $\hat{\bar
S}_{c_q}$ the mapped elements of $E_q$ and $\bar S_{c_q}$ in GF(4),
see Eq.~(\ref{eq:isomorphic1}).} and~$p[x]$ the probability
of~$\bm{u}_{E_c}\cdot \bm{v}_{S_c}=0$ (excluding the $q^{\rm th}$
entry) to take the value~$x$. As $q\in n(c)$,
\begin{equation}
    S_{c_q}\in\{X, Z, Y\}\implies \hat{S}_{c_q} \in \{1, \omega,\bar{\omega}\},x\in \{0, 1, \omega,
    \bar{\omega}\},
\end{equation}
Thus,
\begin{align}
\label{eq:communicatecase1}
    p_q[0]&=\left(p[0]+p[1]\right)/2=p_q[1],    \\
\label{eq:communicatecase2}
    p_q[\omega]&=\left(p[\omega]+p[\bar{\omega}]\right)/2=p_q[\bar{\omega}].
\end{align}

Similarly, if $s_{c}=-1$, then~$\{S_{c}, E_{c}\}=0$, which implies
that either~$\bm u_{E_c} \cdot \bm {v}_{S_c}=\omega$ or~$\bm u_{E_c}
\cdot \bm {v}_{S_c}=\bar{\omega}$ because ${\rm Tr}(\omega)=1$
and~${\rm Tr}(\bar{\omega})=1$. According to
Table~\ref{table:isomorphictable1},
\begin{equation}
\label{eq:communicatecase1}
    p_q[0]=\left(p[\omega]+p[\bar{\omega}]\right)/2=p_q[1],
\end{equation}
\begin{equation}
\label{eq:communicatecase2}
    p_q[\omega]=\left(p[0]+p[1]\right)/2=p_q[\bar{\omega}].
\end{equation}
With this knowledge, the message $m_{c\to q}$ can be computed by:
\begin{equation}
\label{eq:computemessage}
    m_{c \to q}(E_q)=p_q[\hat{E_q}\times
\hat{\bar{S}}_{c_q}].
\end{equation}
Given $m_{c \to q}(E_q)$, $m_{q \to c}(E_q)$ can then be derived
directly by substituting Eq.~(\ref{eq:computemessage}) into
Eq.~(\ref{eq:BPequation2}).

Decoding based on~$GF(4)$ affords the advantage that sparse quantum
codes can be decoded without losing correlations between errors that
would otherwise impact the coding scheme's performance. In
particular sparse quantum codes can be decoded directly regardless
of whether they arise from a classical binary codes or not and
regardless of whether they follow a CSS construction code or not.
For comparison, we henceforth refer to the decoder based on this BP
decoding algorithm as the `standard BP decoder'.

\section{Enhanced feedback BP iterative decoding for sparse quantum codes}
\label{sec:heuristic}
In 2008 PC08 proposed a random perturbation strategy based on the syndrome
of the BP decoder output\footnote{In addition to the random
perturbation technique, PC08 also proposed freezing and collision techniques.
However, their simulations and our own
simulations indicate that the random perturbation technique has the best
performance of the three approaches. Therefore, we adopt the random
perturbation technique as the comparator.} to
overcome the symmetric degeneracy problem.
The random perturbation method is simple and efficient but unfortunately overlooks many
errors. As pointed out in PC08, all errors in their
simulations could be attributed to the decoder rather than to the
finite minimal distance of the code.

We observe the same problems as PC08 do when using the random
perturbation method. In particular, most errors in the simulations
are \emph{detected} errors, not the undetected errors. This result
suggests that feeding back useful information to the standard BP
decode could help. The feedback would help the BP decoder determine
a valid output $E_{\rm out}$, whose syndrome $s(E_{\rm out})$ is
identical with the observed syndrome~$s$. Then $E_{\rm out}$ will
equal~$E$ with high probability, for~$E$ the error occurring on the
transmitted quantum state during transmission. In this section, we
present our enhanced feedback BP iterative decoding algorithm, which
provides useful information to the BP decoder based on exploiting
not only the syndrome but also the stabilizer itself and the channel
model in the feedback procedure.

Then suppose that, subsequent to the standard BP decoding procedure,
$s(E_{\rm out})$ is not identical with the observed syndrome~$s$.
Then evidently at least one entry of~$E_{\rm out}$ occurring on one
of the qubits connected to check~$S_c$ has an error, for $S_c$ a
frustrated check. If we can reset the initial probability
distributions for errors occurring on the qubits connected to $S_c$
in a more reliable way, but not simply by adopting the prior
probability distributions obtained from the channel model, more
useful information could be provided to the BP decoder. Furthermore
symmetric degeneracy could also be overcome as well. Thus it is
reasonable to expect that BP decoding ability will be significantly
improved.

Our new feedback BP iterative decoder is devised as follows. First
the standard BP decoder is used to identify the error~$E$ based on
the observed syndrome and the prior probability distributions of
each entry of~$E$. When the standard BP decoder fails~($s(E_{\rm
out}) \neq s$), we randomly choose a frustrated check, say~$S_c$.
There are only two results for this frustrated check: $s_c=-1$
and~$S_c\cdot E_{\rm out}=1$ or else $s_c=1$ and~$S_c \cdot E_{\rm
out}=-1$. Either way we choose a random qubit that connects to
$S_c$, for example qubit~$q$, and then reset $p_q(E_q)$ according to
the value of~$S_{c_q}$~(the entry to which qubit~$q$ corresponds
in~$S_{c}$) and the channel model as follows.

Let~$\Sigma_{1,2,3}\in\left\{X,Z,Y=XZ\right\}$ be three distinct
elements. Suppose $S_{\rm c_q}=\Sigma_1$, and the channel is the
depolarizing channel with crossover probability~$p=1-p_I$, for~$p_I$
established in Eq.~(\ref{eq:depolarizingchannel}). Then, if~$s_{\rm
c}=-1$ and~$S_{\rm c} \cdot E_{\rm out}=1$, we reset
\begin{equation}
\label{eq:resetting1}
    p_q(I)=(1-p_I)/2=p_q(\Sigma_1),\,
    p_q(\Sigma_2)=p_I/2=p_q(\Sigma_3).
\end{equation}
For example, if~$S_{\rm c_q}=X$, we obtain
\begin{equation}
    p_q(I) = (1-p_I)/2 = p_q(X),\,
    p_q(Z) = p_I/2 = p_q(Y).
\end{equation}
Instead suppose $s_{\rm c}=1$ and~$S_{\rm c} \cdot E_{\rm out}=-1$.
Then we reset
\begin{equation}
\label{eq:resetting2}
    p_q(I)=p_I/2=p_q(\Sigma_1),\,
    p_q(\Sigma_2)=(1-p_I)/2=p_q(\Sigma_3).
\end{equation}
In this case, if~$S_{\rm c_q}=X$, we obtain
\begin{equation}
\label{eq:cq=X}
    p_q(I)=p_I/2=p_q(X),
    p_q(Z) = (1-p_I)/2=p_q(Y).
\end{equation}

Next, following a similar strategy to PC08's, we feed back this
adjustment to the standard BP decoder and let it iterate with this
modified probability distribution for~$T_\texttt{pert}$~(a
pre-determined fixed number of the reiteration) steps. If BP halts
during this period~(i.e.\ $s(E_{\rm out})$ corresponds to the
syndrome~$s$), the procedure is complete and terminates.

If~$c$ is still frustrated, we restore the probability
distribution~$p_q(E_q)$ and choose a different qubit connected to
$c$, say qubit~$q'$, and reset $p_q'(E_q')$ based on the same
strategy that has already been used for resetting~$p_q(E_q)$.
If~$S_{\rm c}$ is not frustrated, but the halting condition is not
yet satisfied, we choose another frustrated check $S_{\rm c'}$ and
adjust the probability distribution of the error occurring on one of
its qubits as we do for a frustrated check $S_{\rm c}$.

Now suppose that the decoder still cannot yield a valid output after
trying a pre-determined number $n_a$ feedback adjustments for qubits
according to the procedures mentioned above. In other words, suppose
that, after traversing $n_a$ entries associated with frustrated
checks, no valid result is yielded by the decoder. In this case, we
allow the decoder to terminate this procedure and report a failure.
Without the termination condition, the decoder is wasting time
traversing all entries associated with frustrated checks and thereby
introducing unwanted redundancy into the procedure.

Eqs.~(\ref{eq:resetting1}) and (\ref{eq:resetting2}) are justified
for the following reason. Empirically, the decoder is naturally too
biased towards the identity~$I$~\cite{PC08} due to the prior
probability distribution obtained from the channel model. This
biased probability distribution will lead to $S_{\rm c}\cdot E_{\rm
out}=1$ while $s_{\rm c}=-1$. Heuristically, we reset $p_q(E_q)$ to
make it anti-commute with $S_{\rm c_q}$ with a greater probability,
since at least one entry of~$E$ anti-commutes with its corresponding
entry in~$S_{c}$. The channel is depolarizing that is Pauli
errors~$X$, $Z$, $Y$ occur with equal probability, so, in
Eq.~(\ref{eq:resetting1}), we let $\Sigma_2$ and~$\Sigma_3$ share
the probability as~$I$.\footnote{Because $(1-p_I)\ll p_I$, we let
$I$ and~$\Sigma_1$ share $(1-p_I)$ equally for simplicity.}

When~$s_{\rm c}=1$, setting $E_{\rm q}$ to~$I$ is conducive to
obtaining~$S_{\rm c}\cdot E_{\rm out}=1$. Another Pauli operator
could also be conducive to making~$S_{\rm c}\cdot E_{\rm out}=1$ but
is ignored by the prior probability distribution. We can readily
ascertain this Pauli operator according to the entries of~$S_{\rm
c}$. According to Eq.~(\ref{eq:resetting2}), we let this Pauli
operator~($\Sigma_1$) share the same probability~$p_I$ with~$I$ and
have~$\Sigma_2$ and~$\Sigma_3$ share the small probability~$1-p_I$
equally.

Our strategy retains PC08's capability of breaking the symmetric
degeneracy while feeding back more useful information to the BP
decoder than PC08 in order to help the decoder determine a valid
output. More useful information comes from retaining full
information about the entire stabilizer rather than just a syndrome.
Specifically, our feedback adjustment strategy is not solely based
on the syndrome but also on the channel model and on the individual
values of the entries of the frustrated checks~(the checks are just
the generators of the stabilizer).

\section{A Case Study}
\label{sec:simpleexample} In this section we illustrate our enhanced
feedback iterative decoding algorithm to show how it succeeds in
helping the BP decoder find a valid output. For $[[n, k; c]]$
denoting an EA QECC that encodes $k$ qubits into $n$ qubits with the
help of~$c$ ancillary ebits~\cite{BDH06}, we construct a simple EA
QECC for~$[[4, 1;1]]$ as follows. According to the EA QECC
formalism, we start with a classical~$[4, 2]$ quaternary code with
check matrix
\begin{equation}
\label{eq:classicalcheckmatrix}
    H_{\rm c}=\left(\begin{array}{cccc}1&\omega&1&0\\1&1&0&1\end{array}\right).
\end{equation}
First we transform $H_{\rm c}$ to
\begin{equation}
 \label{eq:extendedclassicalmatrix}
    H_{\rm c}'
        =\left(\begin{array}{c} H_{\rm c}\\ \omega H_{\rm c} \end{array}\right)
        = \left(\begin{array}{cccc}
            1&\omega &1&0\\ 1&1&0&1\\ \omega &\bar{\omega}&\omega &0\\
            \omega &\omega &0&\omega \end{array}\right),
\end{equation}
then transform $H_{\rm c}'$ to a set of~(perhaps non-commuting)
generators
\begin{equation}
\label{eq:initialstabilizer}
    H_q=\left(\begin{array}{cccc}
            X&Z&X&I\\X&X&I&X\\Z&Y&Z&I\\Z&Z&I&Z
        \end{array}\right).
\end{equation}

Next we transform $H_q$ into a canonical form $\hat{H}_q$ by
multiplying the third generator by the second generator and by
multiplying the fourth generator by the first and second generator.
Thus, we obtain
\begin{equation}
\label{eq:canonicalstabilizer}
    \hat{H}_q
        =\left(\begin{array}{cccc}
            X&Z&X&I\\ X&X&I&X \\Y&Z&Z&X\\Z&X&X&Y
        \end{array}\right).
\end{equation}
Now, it is easy to check that the first generator anti-commutes with
the second generator, and the last two generators commute with each
other as well as commuting with the first two generators.
According to the EA QECC formalism, this coding scheme needs one ebit to assist
encoding. The extended commuting set of generators is
$\{\tilde{H}_q\}$ for
\begin{equation}
\label{eq:extendedcanonicalstabilizer}
    \tilde{H}_q
        =\left(\begin{array}{cccc|c}
            X&Z&X&I&X\\ X&X&I&X&Z\\Y&Z&Z&X&I\\Z&X&X&Y&I
        \end{array}\right),
\end{equation}
which is just the stabilizer of~$[[4, 1; 1]]$.

It is easy to check that $\tilde{H}_q$ is isomorphic to
\begin{equation}
\label{eq:initialcanonicalstabilizer}
    \tilde{H}_q'
        =\left(\begin{array}{cccc|c}
            X&I&I&I&X\\Z&I&I&I&Z\\ I&Z&I&I&I\\ I&I&Z&I&I
        \end{array}\right),
\end{equation}
which means we can encode one logical qubit into four physical bits
with the help of one entangled pair plus two ancilla bits via a
unitary transformation~\cite{BDH06}.
As the half of the entangled pair being held by the receiver is assumed to be
error-free, the error-correcting capacity of~$[[4, 1; 1]]$ only
depends on the left part of~$\tilde{H}_q$, which corresponds to
the four transmitted qubits held by the sender.

\begin{figure}
\noindent\begin{minipage}{\linewidth}
\begin{minipage}[htbp]{.48\linewidth}
\centering\includegraphics[width=1.78in]{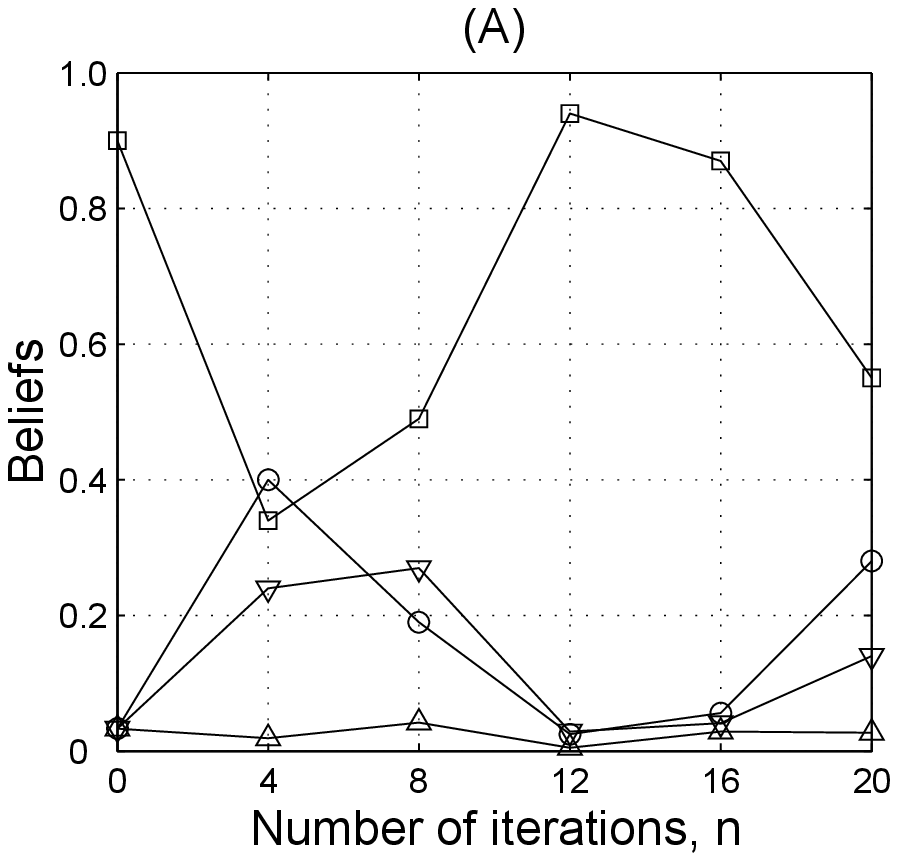}
\end{minipage}
\begin{minipage}[htbp]{.48\linewidth}
\centering\includegraphics[width=1.8in]{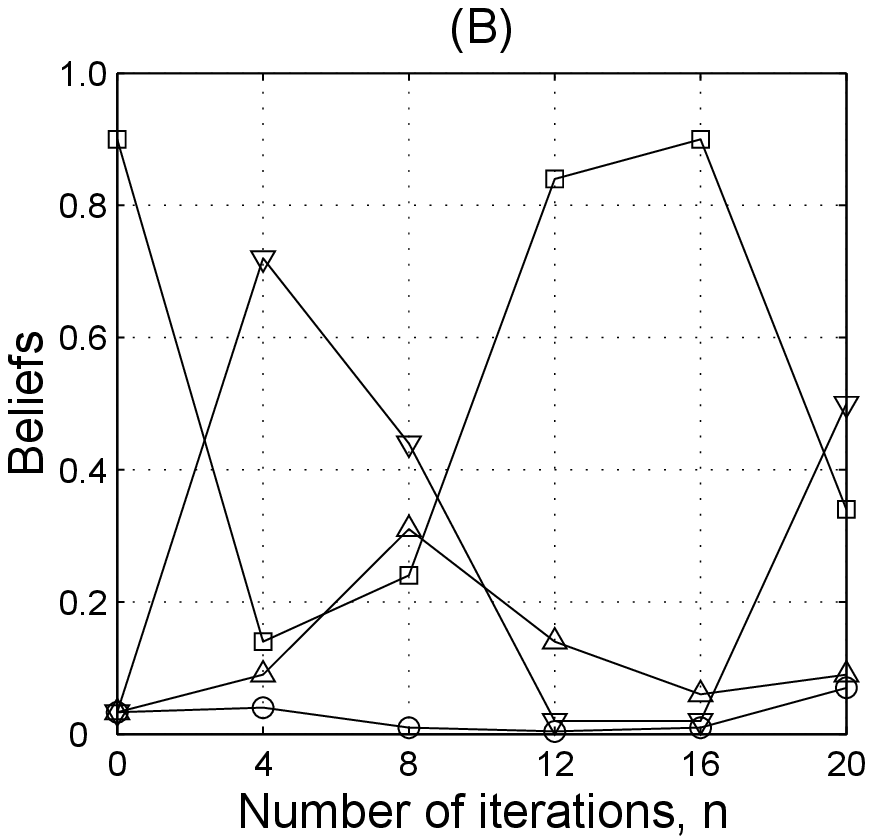}
\end{minipage}
\end{minipage}
\noindent\begin{minipage}{\linewidth}
\begin{minipage}[htbp]{.48\linewidth}
\centering\includegraphics[width=1.78in]{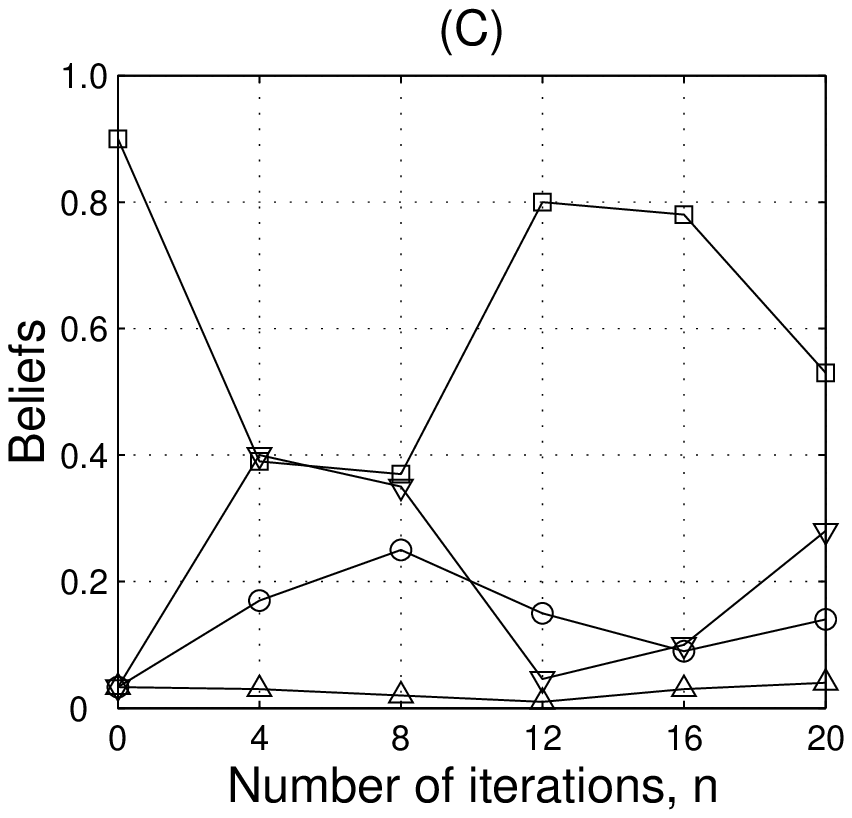}
\end{minipage}
\begin{minipage}[htbp]{.48\linewidth}
\centering\includegraphics[width=1.8in]{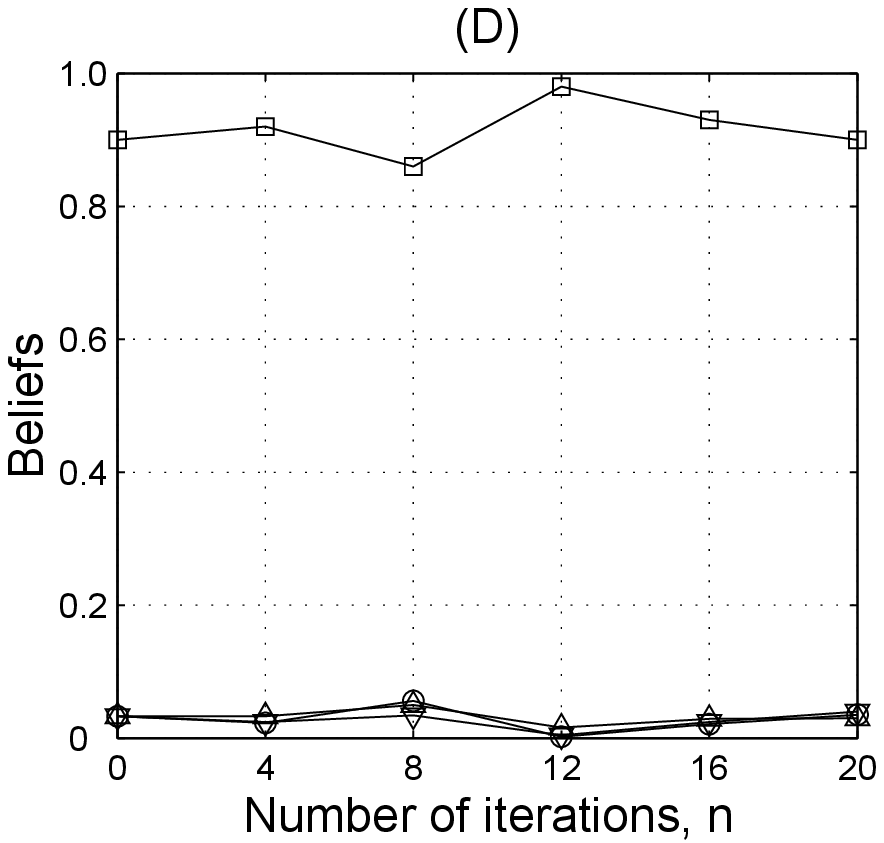}
\end{minipage}
\end{minipage}
\caption{
    Beliefs as a function of number of iterations~$n$ for an EA QECC
    with stabilizer generators $XZXIX$, $XXIXZ$, $YZZXI$ and~$ZXXYI$ and
    with syndrome $(-1,+1,+1,+1)$ using the standard BP decoding
    algorithm. The prior~$p_q(E_q)$ of the four sent qubits are obtained
    from the depolarizing channel with~$p=0.1$. ($A$), ($B$), ($C$) and
    ($D$) show the beliefs for qubits~$1$, $2$, $3$ and~$4$, respectively.
    Qubit~$5$ is assumed to be held by the receiver
    and thus to be error-free.
    The symbols~$\Box$, $\Delta$, $\bigcirc$ and~$\nabla$
    indicate the beliefs for~$I$, $X$, $Z$ and~$Y$, respectively.
    In this case, the output of the standard BP decoder is $E_{\rm out}=IYII$, which is a detected error.
    }
\label{fig:standardBP}
\end{figure}

\begin{figure}
\noindent\begin{minipage}{\linewidth}
\begin{minipage}[t]{.48\linewidth}
\centering\includegraphics[width=1.77in]{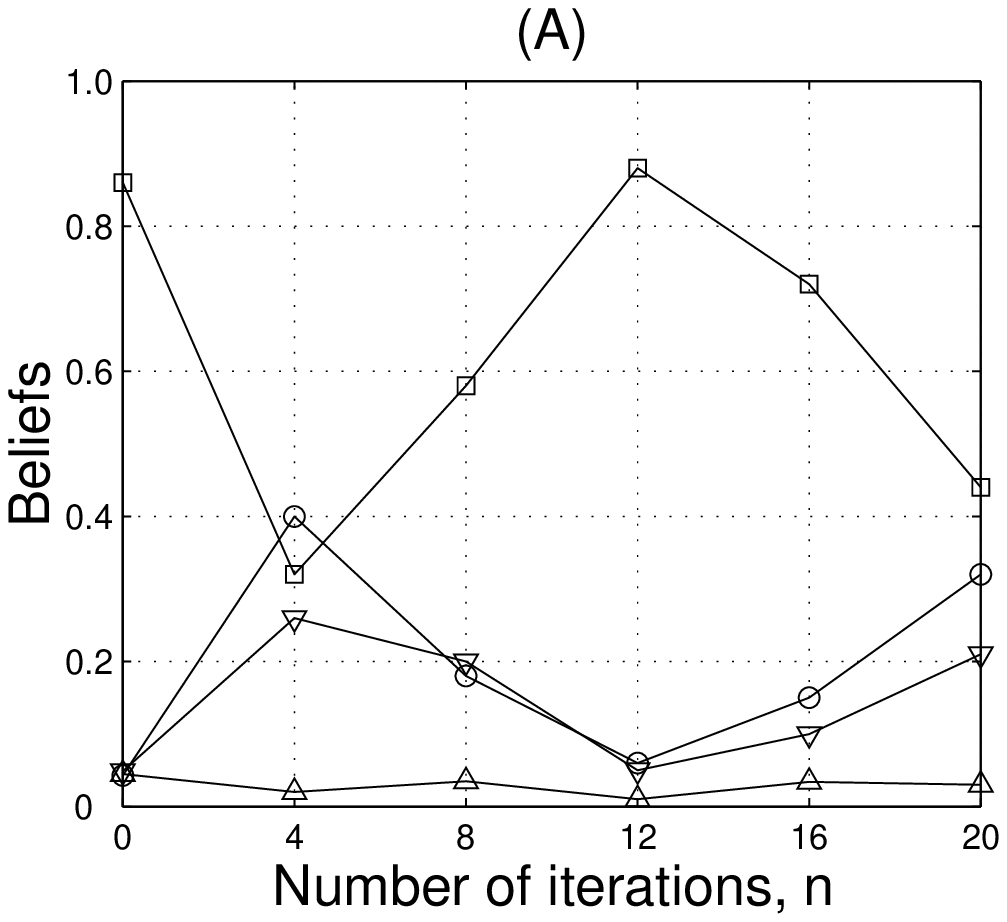}
\end{minipage}
\begin{minipage}[t]{.48\linewidth}
\centering\includegraphics[width=1.8in]{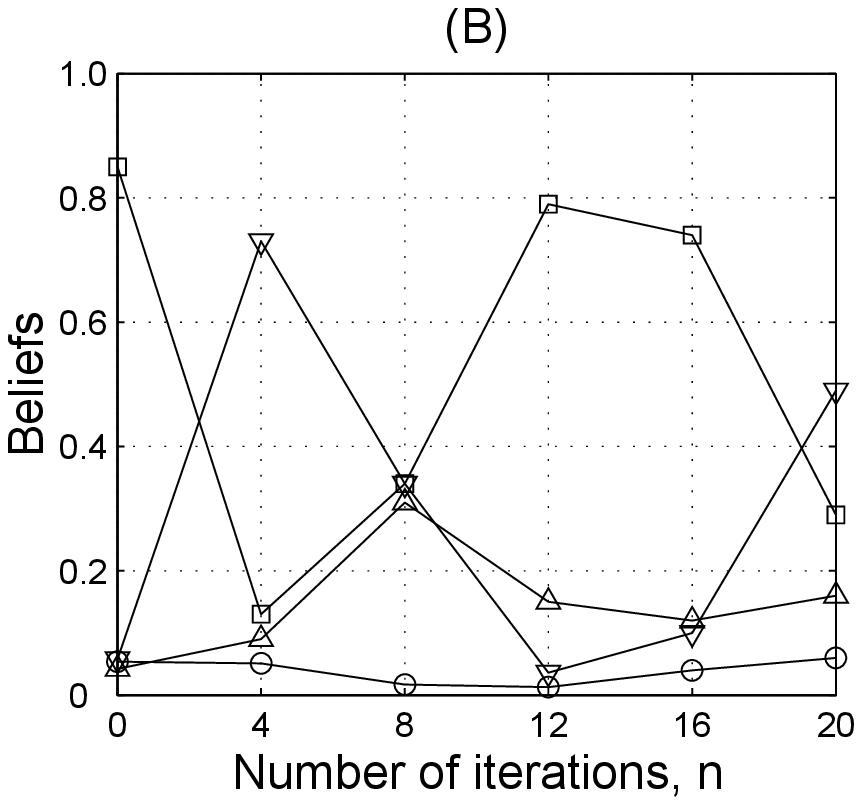}
\end{minipage}
\end{minipage}
\noindent\begin{minipage}{\linewidth}
\begin{minipage}[t]{.48\linewidth}
\centering\includegraphics[width=1.77in]{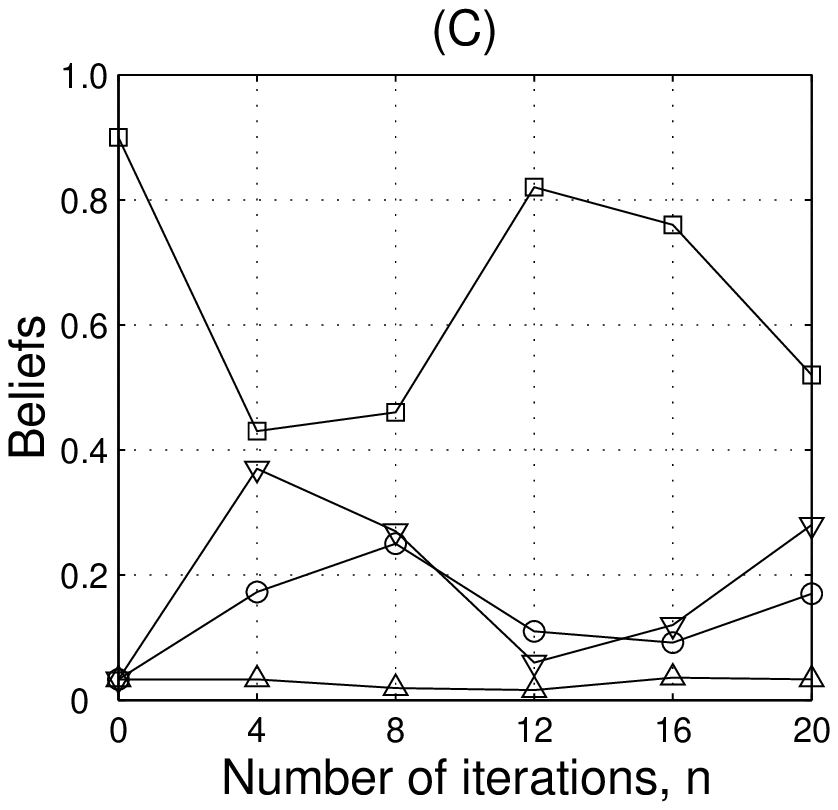}
\end{minipage}
\begin{minipage}[t]{.48\linewidth}
\centering\includegraphics[width=1.8in]{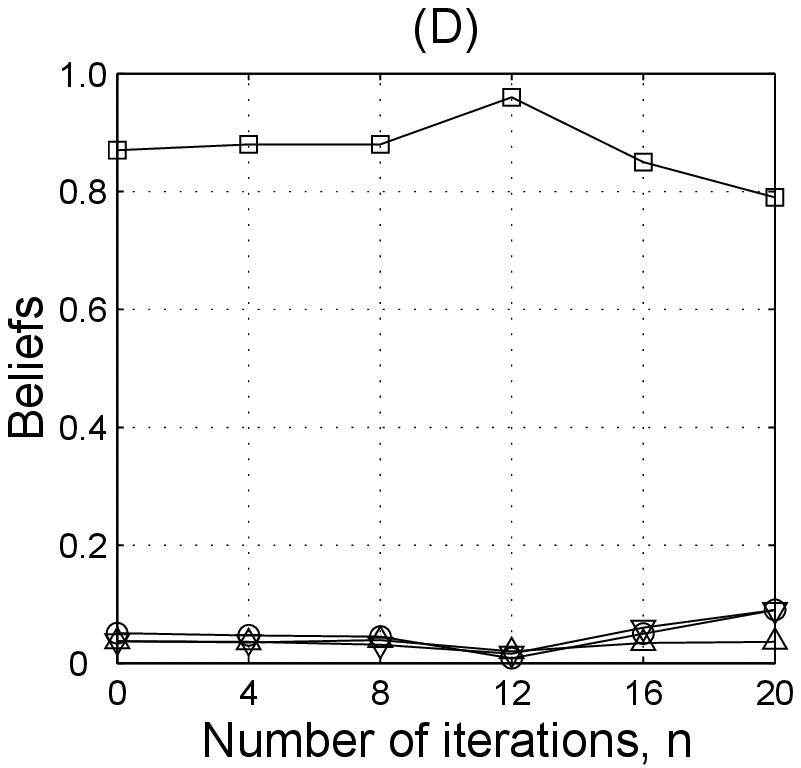}
\end{minipage}
\end{minipage}
\caption{
    Same as for Fig.~\ref{fig:standardBP} but using the standard BP decoding
    algorithm replaced by PC08 and applying random perturbation strength~$\delta=1$ to
    the prior for qubits~$1$, $2$ and~$4$.
    }
\label{fig:syndrome-basedBP}
\end{figure}

Suppose the qubits of~$[[4, 1;1]]$ are sent through a depolarizing
channel and error~$E=IIZX$ is applied to the four transmitted
qubits. Measurement of the stabilizer reveals the error syndrome
$s=(-1,+1,+1,+1)$. In Fig.~\ref{fig:standardBP}, we display the
performance of the standard BP decoder under this setting.
From Fig.~\ref{fig:standardBP}, it is easy to check that the
syndrome of the output of the standard BP decoder is
$(-1,-1,-1,-1)$.\footnote{It is not surprising that the standard BP
decoder has failed because $[[4, 1;1]]$ is designed to correct one
error but not two errors.} We now use the standard BP decoder with
PC08's random perturbation. Comparing between the error syndrome
$s=(-1,+1,+1,+1)$ and the syndrome of the output of the standard BP
decoder $(-1,-1,-1,-1)$, the probability distributions of the errors
occurring on the qubits connected to the second, third or fourth
checks can be reset.  For all these cases, the decoder could not
yield an appropriate recovery. Here as an example, we use PC08's
random perturbation strategy for the frustrated check $S_2$ and show
the corresponding performance in Fig.~\ref{fig:syndrome-basedBP}.

\begin{figure}
\noindent\begin{minipage}{\linewidth}
\begin{minipage}[t]{.48\linewidth}
\centering\includegraphics[width=1.78in]{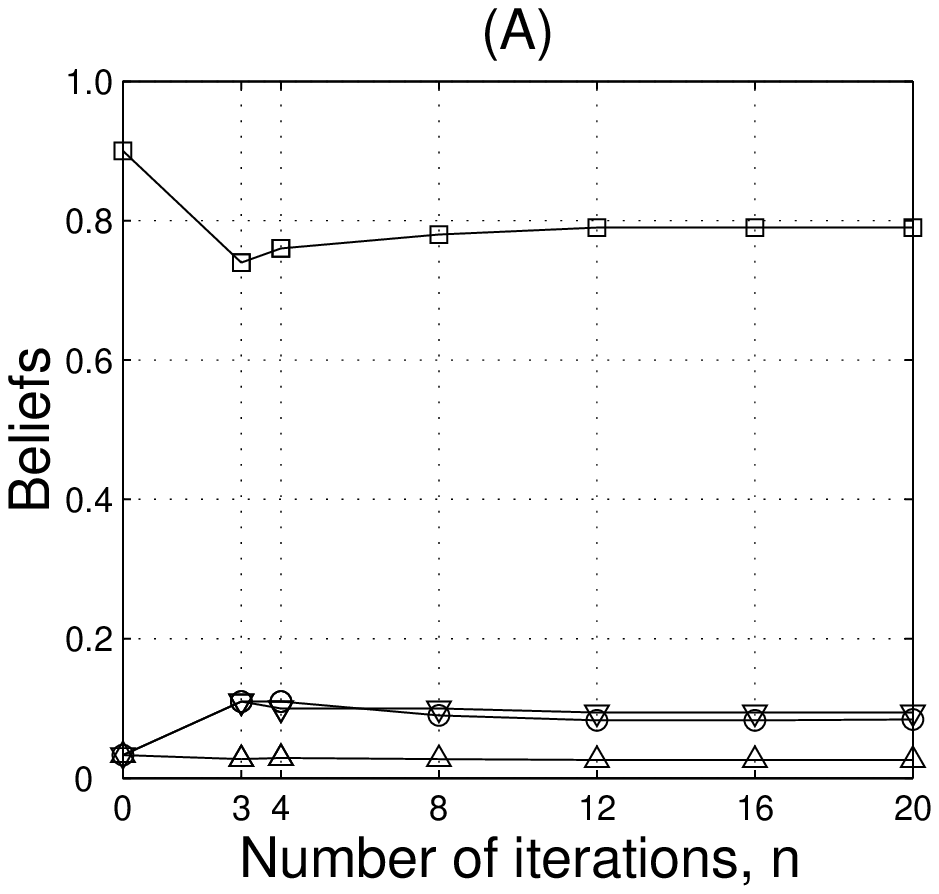}
\end{minipage}
\begin{minipage}[t]{.48\linewidth}
\centering\includegraphics[width=1.8in]{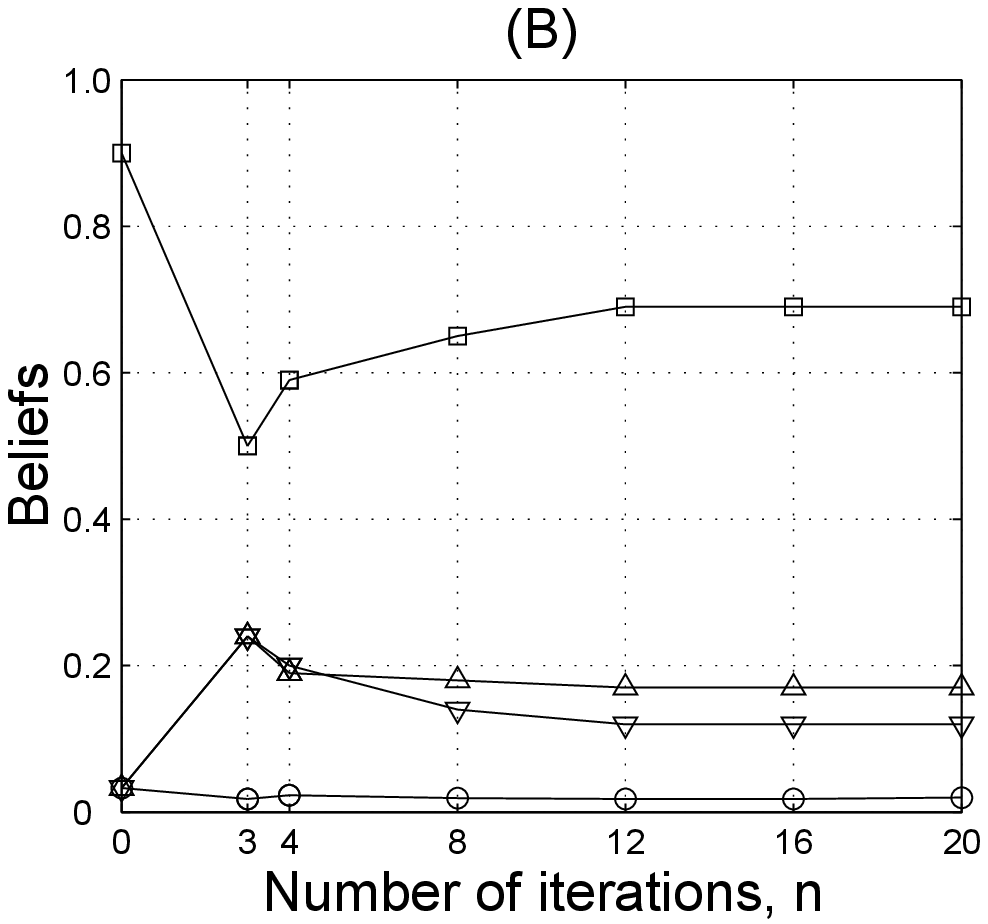}
\end{minipage}
\end{minipage}
\noindent\begin{minipage}{\linewidth}
\begin{minipage}[t]{.48\linewidth}
\centering\includegraphics[width=1.78in]{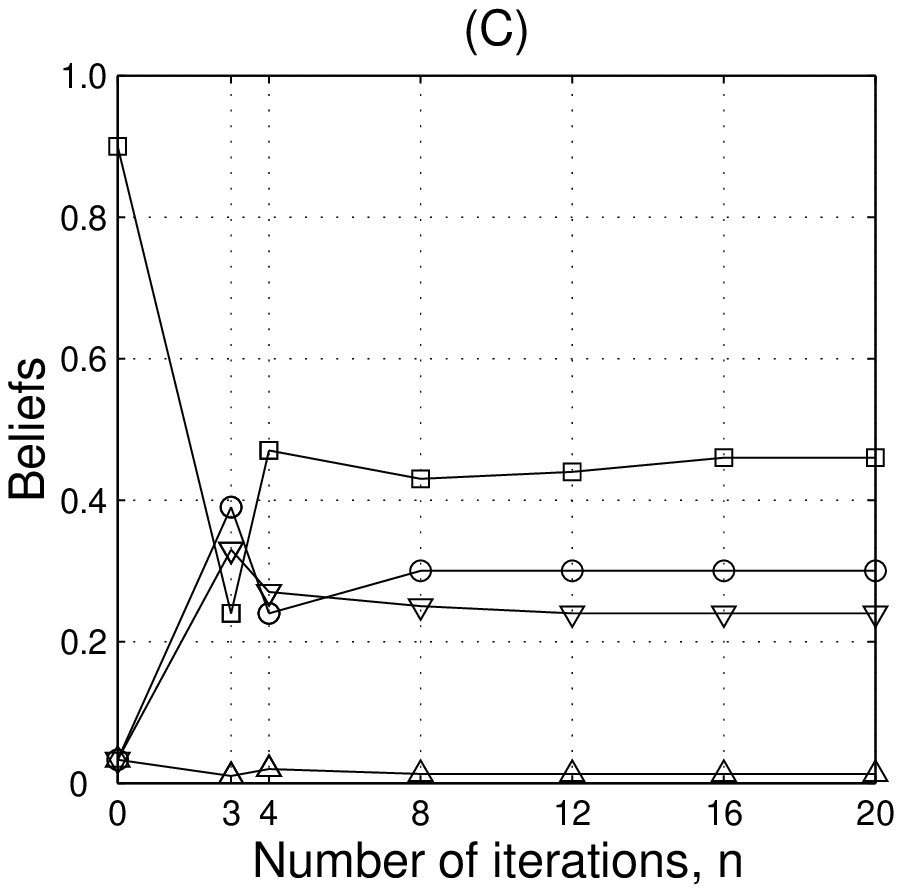}
\end{minipage}
\begin{minipage}[t]{.48\linewidth}
\centering\includegraphics[width=1.8in]{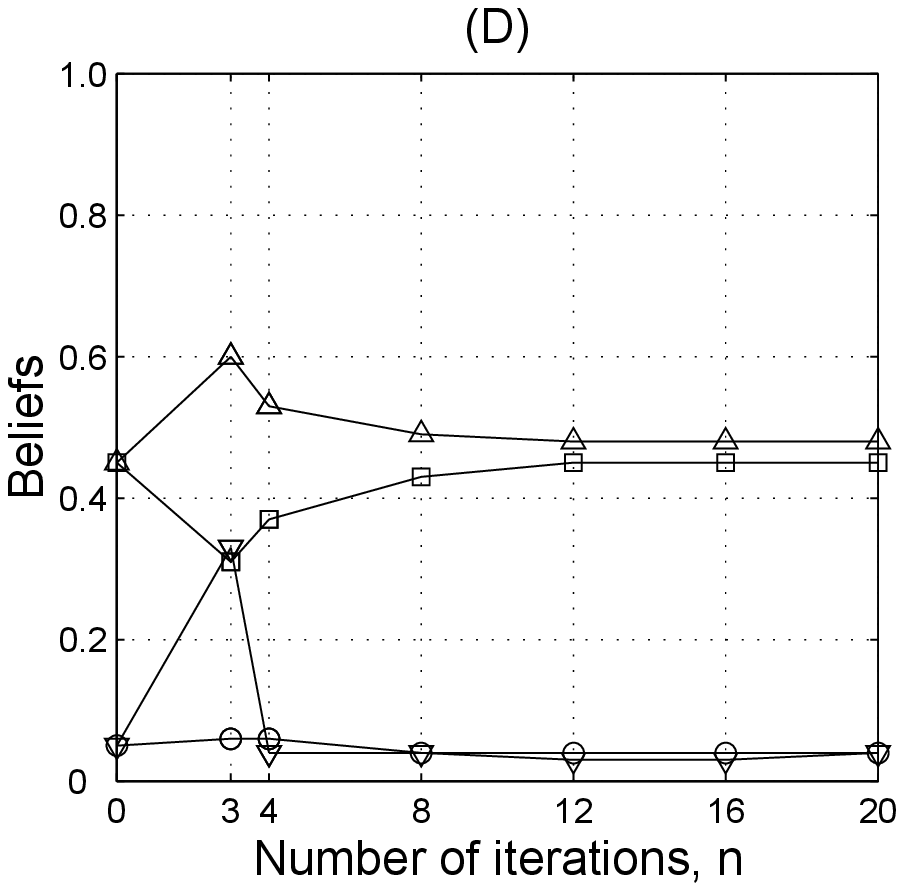}
\end{minipage}
\end{minipage}
\caption{
    Same as for Fig.~\ref{fig:syndrome-basedBP} but using our enhanced feedback iterative decoding algorithm. Here we reset $p_4(E_4)$ according to
    Eq.~(\ref{eq:resetting2}). In this case, only three iterations were required to yield a
    valid output $E_{\rm out}=IIZX$,
    which is exactly the error occurring on [[4, 1;1]] during its transmission.
    }
\label{fig:stabilizer-basedBP}
\end{figure}

Finally, we use our enhanced feedback BP decoding algorithm.
By the same method, our feedback strategy can reset the probability
distributions of the errors occurring on the qubits connected to the
second, third or the fourth checks. When the decoder chooses the
fourth entry ($X$) of the frustrated check $S_2$ and resets
$p_4(E_4)$ according to Eq.~(\ref{eq:resetting2}), the correct
decoding result arises in just a few iterations. We show the
performance of our approach in Fig.~\ref{fig:stabilizer-basedBP}.

In this example, PC08's method could not help the standard BP
decoder to yield a valid output, However, our enhanced feedback BP
decoding algorithm yields a valid output in just a few iterations.
In fact even if the error occurring on the first four qubits is not
$IIZX$ but $YZII$, which has the same syndrome $s=(-1,+1,+1,+1)$ as
$IIZX$, our decoding output $IIZX$ can still recover the transmitted
quantum state because
\begin{equation}
    IIZXI\times YZIII=YZZXI,
\end{equation}
which is just the third generator of the extended stabilizer.
Because it is hard to check whether~$E\cdot E_{\rm out}\in S$ when
the number of generators of~$S$ is large, we choose $E=E_{\rm out}$
as the success criterion of the decoding result in our simulations.

\section{Simulation Results}
\label{sec:simulation} We have applied our enhanced feedback
iterative BP decoding algorithm to a variety of sparse quantum
codes, including conventional sparse quantum codes and EA sparse
quantum codes over depolarizing channels. In each case, our improved
BP decoder yields significantly lower BER over both the standard BP
decoder and the BP decoder with PC08's random perturbation. In the
following subsections, we simulate decoding of sparse quantum codes
with different code parameters~(including block lengths, rates, and
row weight) under the three decoders to demonstrate the superiority
of our approach.

\subsection{Example: Conventional Sparse Quantum Codes}
\label{subsec:conventional}
Conventional sparse quantum codes can be constructed from sparse classical dual-containing codes.
One of the most successful dual-containing constructions is the so-called ``Construction B''~\cite{MMM04},
which is built as follows.
First we take an $n/2 \times n/2$ cyclic matrix $C$ with row weight $L/2$, and define
\begin{equation}
    H_0=\left[C, C^T\right].
\end{equation}
Then we delete some rows from $H_0$ to obtain a matrix $H$ with~$m$ rows.
By construction, $H$ is dual-containing.
A conventional sparse quantum code with length~$n$ can thus be given according to CSS construction.

\begin{figure}[htbp]
\centering
   \includegraphics[width=7.5cm]{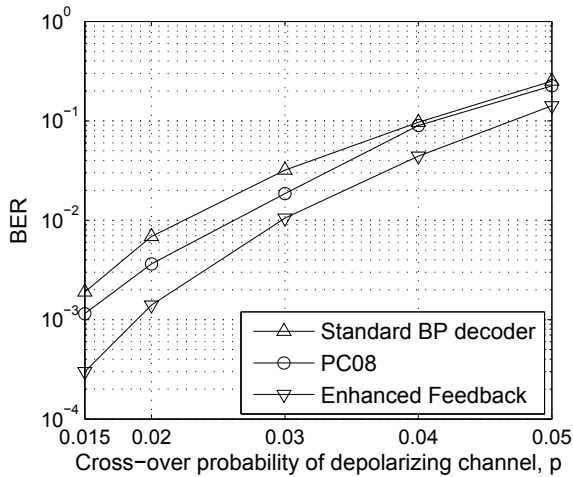}
\caption{
    Block error rate vs cross-over probability of
    depolarizing channel~$p$ for three different BP decoders when applied to Conventional.
    The maximum number of iterations is $90$. The maximal number of
    iterations between each perturbation is $40$. The strength of the
    random perturbation used here is $\delta=0.1$. The maximal number of entries
    traversed by enhanced feedback decoder round down to one fifth of the code length,~which equals 25 here.}
    \label{fig:conventionalexample}
    \end{figure}

Our first example is based on this strategy. We first construct a
cyclic binary LDPC code [63, 37] with row weight~$8$ based on finite
geometries~\cite{KLF01}. This code has a $63 \times 63$ cyclic check
matrix with~$26$ independent rows and~$37$ redundant rows. Hence, we
can construct a conventional sparse quantum code with~$n=126$,
$m=26$, and $L=16$. We refer to this $[[126, 74]]$ quantum LDPC code
with rate~$k/n=74/126$ and row weight $16$ as
`Conventional".\footnote{Note that in order to improve the
performance of this code, we have preserved those redundant rows.}
We show the performance of the standard BP decoder, the BP decoder
with PC08's random perturbation and our enhanced feedback BP decoder
when applied to Conventional over depolarizing channels in
Fig.~\ref{fig:conventionalexample}.\footnote{In our simulation, we
take the maximum number of entries
    traversed by the enhanced feedback decoder round down to one fifth of the code length for short codes, one tenth of the
    code length for medium codes and one fortieth for long codes.}
\subsection{Examples: EA Sparse Quantum Codes}
EA sparse quantum codes can be constructed from arbitrary binary or
quaternary sparse classical codes~\cite{BDH06}. Here we present the
decoding of three EA sparse quantum codes that are constructed in
the same way as the construction of~$[[4,1;1]]$ discussed in
Sec.~\ref{sec:simpleexample}. These three codes have different net
rate~(including positive, zero and negative net rates), different
block lengths and different row weight distributions.

\begin{figure}[htbp]
\centering
   \includegraphics[width=7.5cm]{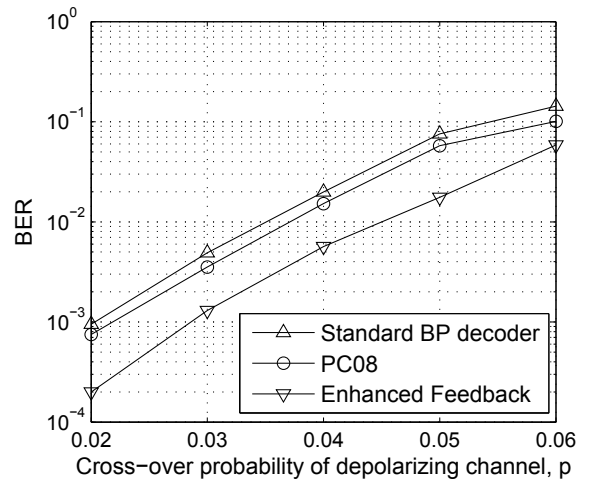}
\caption{Same as for Fig.~4 but replacing Conventional by EA-1.}
\label{fig:EAexample1}
\end{figure}

The first EA sparse quantum code we consider is $[[128, 79; 49]]$,
which is constructed from an irregular quaternary sparse classical
code $[128, 79]$ with girth~$g\geq 6$. The classical code~$[128,
79]$ is designed by using the progressive-edge-growth~(PEG)
algorithm under the constraint that the degree distribution of bit
nodes is~\cite{YBZW09}
\begin{equation}
    \{2:33\%,3:33\%,4:22\%,5:22\%\}.
\end{equation}
As there are $49$ pairs of anti-commuting generators in~$\hat{H}_q$
according to the construction method of Eq.~(\ref{eq:canonicalstabilizer}),
then the EA formalism yields the EA sparse quantum code
\begin{equation}
    [[n, 2k-n+c; c]]=[[128, 2\times 79-128+49; 49]],
\end{equation}
which can be constructed from the classical quaternary code $[n, k]=[128, 79]$~\cite{BDH06}.
The code $[[128,79;49]]$ has a
positive net rate
\begin{equation}
    \frac{k-c}{n}=\frac{79-49}{128}.
\end{equation}
Here we refer to this code as ``EA-1" and show the performance of
the three decoders with respect to EA-1 over depolarizing channels.
The performance, in terms of block error rate, is summarized in
Fig.~\ref{fig:EAexample1}.

Our second EA sparse quantum code is $[[816, 404; 404]]$, which is
constructed from a binary regular sparse classical code $[816,408]$.
The classical~$[816, 408]$, which has a rate~$0.5$, row weight~$6$
and column weight~$3$, was constructed by MacKay~\cite{Maccode1}. As
there are $c=404$ pairs of anti-commuting generators and~$8$
commuting generators in~$\hat{H}_q$, we obtain an EA sparse quantum
code $[[816, 404; 404]]$ according to the EA stabilizer formalism.
Here we take this code as our second example of EA sparse quantum
codes whose net rate is
\begin{equation}
    \frac{k-c}{n}=\frac{404-404}{816}=0,
\end{equation}
and call it ``EA-2". We show the performance of the three decoders
when applied to EA-2 over depolarizing channels in
Fig.~\ref{fig:EAexample2} in terms of BER.\footnote{Net rates for EA
quantum codes can be negative or zero. See Appendix for more
detail.}
\begin{figure}[htbp]
\centering
   \includegraphics[width=7.5cm]{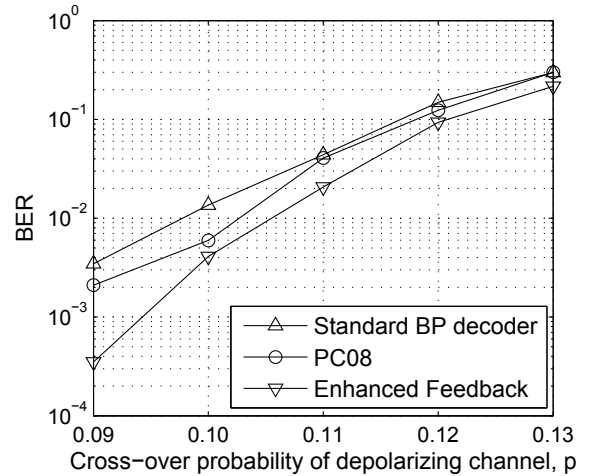}
\caption{Same as for Fig.~4 but replacing Conventional by EA-2 and
changing the maximum number of entries traversed by the enhanced
feedback decoder round down to one tenth of the code length, which
equals 81 here.} \label{fig:EAexample2}
\end{figure}

\begin{figure}[htbp]
\centering
   \includegraphics[width=7.5cm]{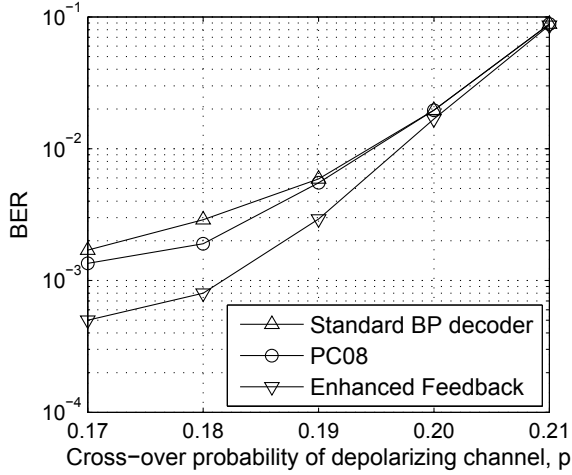}
\caption{Same as for Fig.~4 but using Conventional replaced by EA-3
and changing the maximal number of entries traversed by enhanced
feedback decoder round down to one fortieth of the code length that
is 43.} \label{fig:EAexample3}\end{figure}

The third EA sparse quantum code is $[[1920, 638; 1278]]$, which is
constructed from $[1920, 640]$, a famous sparse classical code
proposed as the international standard for cellular telephones. The
$[1920, 640]$ code, which has a rate $1/3$ and row weight $4$, was
also constructed by MacKay~\cite{Maccode2}. Following the
construction procedure mentioned in Sec.~\ref{sec:simpleexample},
the number of anti-commuting pairs in $\hat{H}_q$ is $1278$; thus,
we obtain an EA sparse quantum code $[[1920, 638; 1278]]$. Here we
take this code as our third example of EA sparse quantum codes, this
time with a negative net rate
\begin{equation}
    \frac{k-c}{n}=\frac{638-1278}{1920}=-\frac{1}{3}.
\end{equation}
We name it ``EA-3" and show the performance of the three BP decoders
when applied to EA-3 over depolarizing channels in
Fig.~\ref{fig:EAexample3}.
\subsection{Time efficiency simulation}
We evaluate and compare the time consumed by each decoder according
to the average number of iterations
\begin{equation}
\text{ANoI}:= \frac{n_t}{n_b},
\end{equation}
consumed by one codeword, with $n_t$ the total number of iterations
to finish decoding $n_b$ blocks. In our simulation $n_b=20000$. In
Figs.~8 and~9 we show the time efficiency (complexity) comparison
for the three decoders applied to Conventional, EA-1, EA-2 and EA-3.
Evidently our algorithm has a smaller ANoI than PC08, which shows
improved decoding efficiency when compared to PC08.
\begin{figure}[htbp]
\centering
   \includegraphics[width=7.5cm]{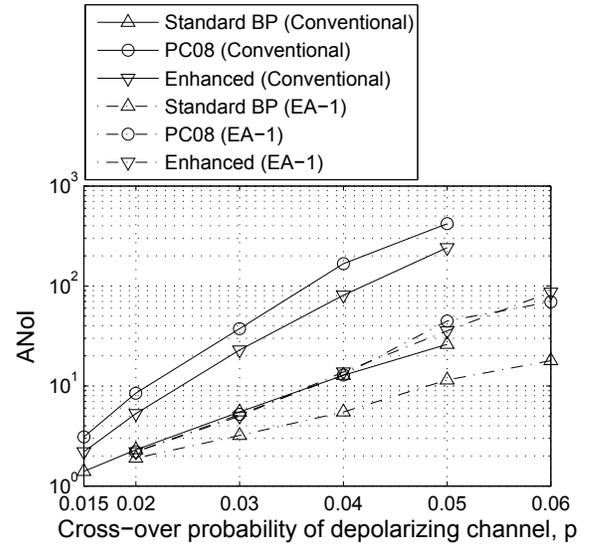}
\caption{Same as for Fig.~4 but replacing block error rate by
average number of iterations and adding EA-1 as another example.}
\label{fig:Averageiteration1}
\end{figure}
\begin{figure}[htbp]
\centering
   \includegraphics[width=7.5cm]{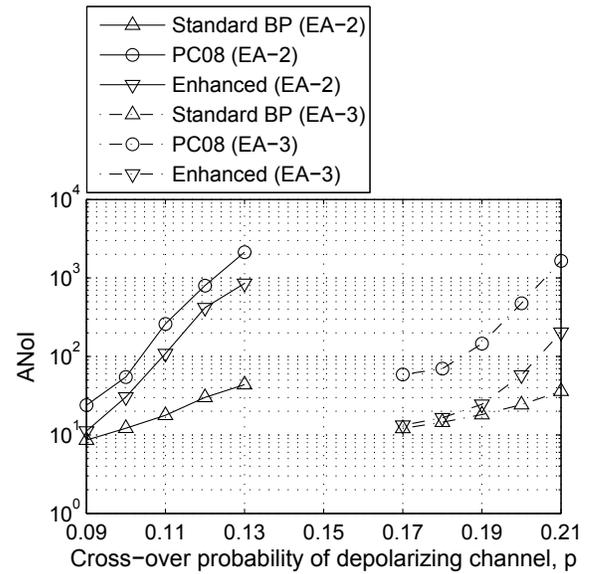}
\caption{Same as for Fig.~8 but replacing Conventional and EA-1 by
EA-2 and EA-3 and changing the maximal number of entries traversed
by enhanced feedback decoder from rounding down to one fifth of the
code length to one tenth of the code length for EA-2 and to one
fortieth of the code length for EA-3.}\label{fig:Averageiteration2}
\end{figure}

\subsection{Discussion}
From
Figs.~\ref{fig:conventionalexample}-\ref{fig:Averageiteration2}, it
is evident that our strategy yields a significantly lower BER with a
lower ANoI for both conventional sparse quantum codes and EA sparse
quantum codes. As an example of the effectiveness of our new
decoder, consider the case~$p\approx 0.015$, Fig.~4 shows that the
enhanced-feedback BP decoder yields a $6$ dB gain over PC08 and an
$8$ dB gain over the standard BP decoder while reducing the ANoI
from 3.1 for PC08 to 2.2 for our enhanced-feedback BP decoder as
seen in Fig. 8.

For another point of view, in Fig.~\ref{fig:EAexample1}, we can see
that to guarantee EA-1 having a decoding error below $10^{-3}$, the
standard BP decoder allows $p\approx 0.020$ and PC08
allows~$p\approx 0.022$. This value is increased dramatically to
$p\approx 0.029$ using our enhanced feedback BP decoder with a
similar ANoI performance as PC08. As for the decoding of
EA-2~(Fig.~\ref{fig:EAexample2}), when~$p=0.09$, the BER performance
of PC08's method is $1.75$ times better than that of the standard BP
decoder with $2.78$ times more ANoI, and the BER performance of our
approach is approximately~$10$ times better than that of the
standard BP decoder with only $1.3$ times more ANoI.

Notably, interesting results appear in the decoding of EA-3. From
Fig.~\ref{fig:EAexample3} and~\ref{fig:Averageiteration2} we see
that PC08 could not significantly improve the standard BP decoder
even with high ANoI, yet our approach does improve the performance
of EA-3 significantly with almost the same ANoI as the standard BP
decoder excepting the last two nodes.

Our algorithm is capable of turning the detectable errors into new
outputs whose syndromes are identical with corresponding observed
syndromes. We refer to these outputs as \emph{valid} outputs can be
classified into three cases: (1) $E_{\rm out}=E$; (2) $E_{\rm
out}\neq E$ but $E_{\rm out}\times E \in S$; (3) $E_{\rm out}\times
E \notin S$. The first two cases will yield correct outputs that can
recover the sent quantum states successfully. However, it is hard to
check whether $E_{\rm out}\times E \in S$ which makes distinguishing
cases~(2) and~(3) difficult. Therefore, we choose (1) as the success
criterion for the decoding result in our numerical simulation.

\begin{figure}[htbp]
\centering
   \includegraphics[width=7.5cm]{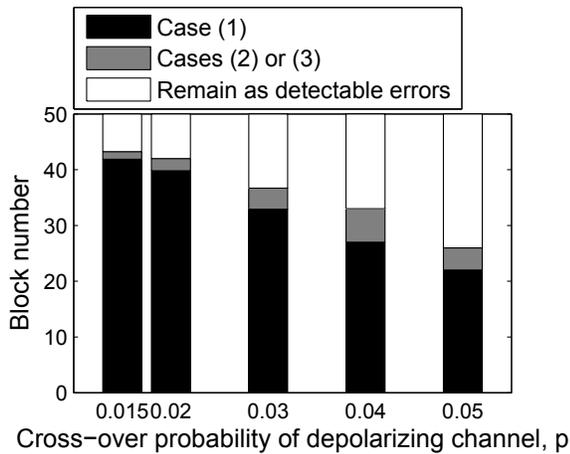}
\caption{Block numbers fall into the three cases mentioned in
context vs cross-over probability of depolarizing channel~$p$ for
our enhanced
    BP decoder when applied to Conventional while the original BP decoder fails 50 times.
    The maximum number of iterations is $90$. The maximal number of
    iterations between each perturbation is $40$. The maximal number of entries
    traversed by enhanced feedback decoder round down to one fifth of the code length.}.
\label{fig:Conventional-proportion}\end{figure}

Some detectable errors are turned into undetectable errors by our
decoder. In order to show how often our algorithm results in an
undetectable error, we classify all the results yielded by our
algorithm (not just valid outputs) when the original BP protocol
fails\footnote{The `fails' means that the syndrome of the output of
the original BP decoder is not identical with the observed syndrome.
In other words, a detected errors yielded by the original BP
decoder.} into three cases: (1); cases (2) or (3); remain as
detectable errors. In Fig.~\ref{fig:Conventional-proportion}, we
take Conventional as an example to depict the number of blocks that
fall into each whereas the original BP protocol fails 50 times.
  It is evident to
see from Fig.~\ref{fig:Conventional-proportion} that most of the
detectable errors will be turned to correct output but not
undetectable errors by using our enhanced feedback BP decoder.

\section{Conclusion}
We have developed an enhanced feedback BP decoding algorithm whose
feedback adjustment strategy is not based solely on the syndrome but
also on the channel model and on the individual values of the
entries of the frustrated checks. Our approach retains the
capability of breaking the symmetric degeneracy while also feeding
back extra useful information to the BP decoder. Therefore, our
feedback adjustment strategy yields a better error correcting
capacity with relative less iterations compared to the proposed
decoding algorithms for sparse quantum codes. We have considered
three cases: the standard BP decoder adapted from classical decoding
to decode quantum codes, the superior BP decoder with Poulin and
Chung's random perturbation~\cite{PC08}, and the BP decoder with our
new feedback adjustment strategy introduced here.

We used the $GF(4)$ representation of quantum codes to construct our
feedback-based algorithm and exploit not just the syndrome but all
the measurement resulting from stabilizer measurements. Then we used
the block error rate~(BER) and average number of iterations (ANoI)
to demonstrate a dramatic better error correcting capacity
improvement with relative less iterations. This result was achieved
using new feedback adjustment strategy vs the two alternatives:
standard BP decoder and the BP decoder with random perturbation.

\appendices
\section*{Appendix}
As shown in Section~V, the net rate for EA-2 and EA-3 are 0 and
$-1/3$ respectively. As pointed out by Brun, Devetak and Hsieh, ``in
general, net rates for EA quantum codes can be positive, negative,
or zero. The zero rate does not mean that no qubits are transmitted
by this code! Rather, it implies that a number of bits of
entanglement is needed that is equal to the number of bits
transmitted." Compared to quantum codes with high net rates, EA
quantum codes with zero or negative net rates employ many more
physical qubits including a great number of bits of entanglement
that being assumed to be error-free,\footnote{The part of ebits held
by the receiver is assumed to be not subject to errors according to
usual practices.} thus these codes can tolerate stronger noise.
Therefore, it is not surprising that EA-2 and EA-3 greatly increase
$p$ the cross-over probability of depolarizing channel when compared
to quantum codes with similar code length and higher net rates.

\begin{figure}[htbp]
\centering
   \includegraphics[width=7.5cm]{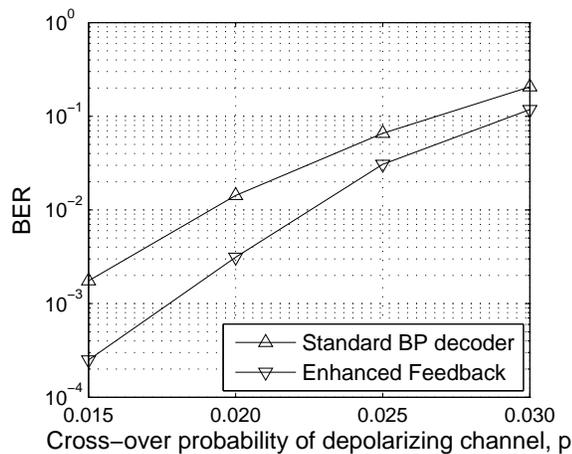}
\\
\caption{Block error probability vs cross-over probability of
    depolarizing channel~$p$ for the standard BP decoder and for our
    algorithm when applied to EX. The maximum number of iterations
    is $90$. Both the number of
    iterations between each perturbation and the maximal entries
    traversed by our decoder are $40$.} \label{fig:EX}
\end{figure}

For comparison, based on ``Construction B"~\cite{MMM04} (same as the
construction of Conventional), we construct a quantum code with a
similar code length as EA-2 but having a relative higher net rate.
We name this example as EX which has a code length of 800 and a net
rate $1/2$. In Fig.~\ref{fig:EX}, we see that EX has comparable BER
performance to other examples presented in existing literatures
which have similar parameters as EX but significantly worse than
EA-2.\footnote{Please refer to~\cite{PC08} and~\cite{HYH}. We can
see that the BER performance of EX is similar with the example
presented in~\cite{PC08} and a little worse than the example
presented in~\cite{HYH} by using standard BP decoder.}

\section*{Acknowledgment}
The authors appreciate critical comments by D. Poulin on an early
version of the manuscript. YJW appreciates financial support by the
China Scholarship Council. BCS received financial support from
\emph{i}CORE, and BMB and XMW received financial support from the
973 Project of China under Grant No. 2010CB328300, the
NSFC-Guangdong Jointly Funded Project of China under Grant No.
U0635003, and the 111 Program of China under Grant No. B08038. BCS
is supported by a CIFAR Fellowship. \ifCLASSOPTIONcaptionsoff
  \newpage
\fi

\end{document}